\documentclass{nature}

\usepackage{graphicx}
\usepackage{subfigure}
\usepackage{amsmath}
\usepackage{amssymb}

\newcommand{\aftr}{ }

\title{Using the transit of Venus to probe the upper planetary atmosphere}

\author{Fabio Reale$^{1,2}$,
Angelo F. Gambino$^{1}$,
Giuseppina Micela$^{2}$,
Antonio Maggio$^{2}$,
Thomas Widemann$^3$,
Giuseppe Piccioni$^4$
}

\begin{document}

\maketitle

\begin{affiliations}
\item Dipartimento di Fisica e Chimica, Universit\`a di Palermo, Piazza del Parlamento 1, 90134 Palermo, Italy
\item INAF/Osservatorio Astronomico di Palermo, Piazza del Parlamento 1, 90134 Palermo, Italy
\item Universit\`e de Versailles-Saint-Quentin - ESR/DYPAC EA 2449, Observatoire de Paris - LESIA - UMR CNRS 8109 5, place Jules-Janssen, 92190 Meudon, France
\item INAF-IAPS (Istituto di Astrofisica e Planetologia Spaziali), via del Fosso del Cavaliere 100, 00133 Roma, Italy
\end{affiliations}

\begin{abstract}
The atmosphere of a transiting planet shields the stellar radiation providing us with a powerful method to estimate its size and density. In particular, \aftr{because of their high ionization energy, atoms with high atomic number (Z)} absorb short-wavelength radiation in the upper atmosphere, undetectable with observations in visible light. One implication is that the planet should appear larger during a primary transit observed in high energy bands than in the optical band. The last 
Venus transit in 2012 offered a unique opportunity to study this effect. The transit has been monitored by solar space observations from Hinode and Solar Dynamics Observatory (SDO). We measure the radius of Venus during the transit in three different bands with subpixel accuracy: optical (4500\AA), UV (1600\AA, 1700\AA), Extreme UltraViolet (EUV, 171-335\AA) and soft X-rays ($\sim 10$\AA). We find that, while the VenusÕ optical radius is about 80 km larger than the solid body radius (the expected opacity mainly due to clouds and haze), the radius increases further by more than 70 km in the EUV and soft X-rays. These measurements mark the densest ion layers of Venus' ionosphere, providing information about the column density of CO$_2$ and CO. They are useful for planning missions in situ to estimate the dynamical pressure from the environment, 
and can be employed as a benchmark case for observations with future missions, such as the ESA Athena, which will be sensitive enough to detect transits of exoplanets in high-energy bands.
\end{abstract}


Transits of Venus \cite{transit_venus} are among the rarest of predictable astronomical phenomena.
They \aftr{were} once of great scientific importance as they were used to gain the first realistic estimates of the size of the Solar System in the 18-19th century \cite{Pasachoff2011a}.
The transit of Venus in 2012 was the last one of the 21st century.
Here we describe how observations of this last transit made by space missions
at UV, EUV and X-ray wavelengths give us new insight on the upper atmosphere of the planet.

The X-rays and EUV solar radiation is stopped in the ionosphere of rocky planets around the peak of the ion/electron density. 
Detailed models of the ionosphere structure with the altitude and the solar zenith angle have been developed for Venus \cite{Chen1978a,Nagy1980a,Cravens1981a,Fox2001a,Fox2007a,Fox2011a}, and a radio measurement of the electronic density has been obtained \cite{Patzold2007a}. The photo-ionization induced by the solar EUV radiation produces CO$_2^+$ ions, that are transformed
by photochemistry to O$_2^+$, by reaction with the O \cite{Fox2001a,Witasse2006a}. Models predict that the peak density of CO$_2^+$ and other molecular ions should be around 150-180 km \cite{Fox2011a}.
\aftr{The mass spectrometer onboard the Pioneer Venus Orbiter measured the ion
composition in situ down to periapsis near 150 km, including CO$_2^+$ \cite{Miller1984a}. Despite at the limit of the observation, the peak at low solar zenith angles
appeared to be slightly below 150 km and it should be reasonably similar near the
terminator.}


The knowledge of the density in the ionosphere and neutral atmosphere is very
important for planning the minimum altitude of the fly-by with spacecrafts and entry
probes of Venus, since they are subject to dynamical pressure and also electrical
charging from the environment. \aftr{For example, on June 2014, ESA planned for the first
time an aerobraking with the Venus Express spacecraft which flew down to a minimum
altitude of 129.2 km over the mean surface of Venus, yielding a maximum dynamic pressure of more than 0.75 N/m$^2$ \cite{Svedhem2014a}.}

\begin{figure}[htbp]               
 \centering
   {\includegraphics[width=16cm]{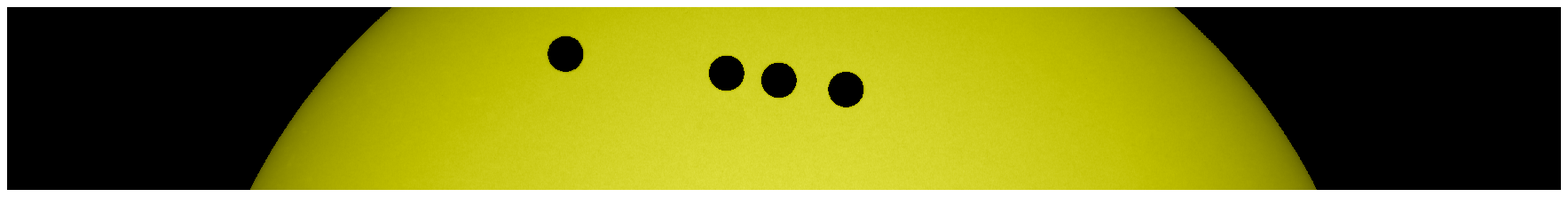}}
   {\includegraphics[width=16cm]{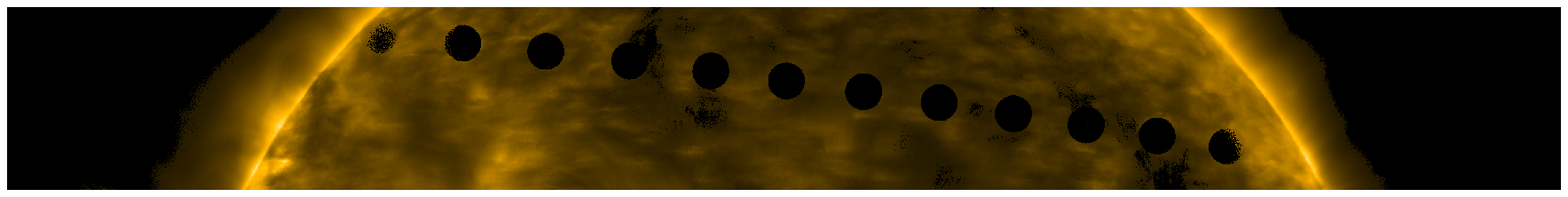}}
\caption{\footnotesize Path of the Venus' transit across the solar disk in the optical (4500 \AA) and EUV (171 \AA) band of the Atmospheric Imaging Assembly on board the Solar Dynamics Observatory. The planet disks mark the range of the selected data.   }
\label{fig:transit}
\end{figure}

The Venus transit started on 5 June 2012 at 22:25 UTC and ended on 6 June 2012 at 04:16 UTC (third contact), during a period of moderate solar activity. It has been observed in great detail by space-borne solar observatories, and in particular by imaging instruments on-board Hinode \cite{Kosugi2007a} and SDO \cite{Pesnell2012a} (Fig.~\ref{fig:transit}). We analysed the observations of the SDO/Atmospheric Imaging Assembly (AIA) \cite{Lemen2012a} in the optical (4500\text{\AA}), UV (1600, 1700\text{\AA}) and EUV (171, 193, 211, 304 and 335\text{\AA}) channels, and of the Hinode/X-Ray Telescope (XRT) \cite{Golub2007a} in the Ti-poly filter that has the maximum sensitivity at $\sim 10$\text{\AA} (Table~\ref{tab:radius}). The AIA and XRT plate scales are 0.60000 arcsec/pixel, and 1.0286 arcsec/pixel, respectively.

We measured the planet radius using the radial intensity profiles from the planet disk center in each selected image (see Supplementary Information). To improve on accuracy and robustness, we derived the intensity values as latitudinal averages, on annuli centered on the planet disk center. \aftr{This allows us to achieve subpixel sensitivity:} the thickness of the concentric annuli is 0.2 pixels (25.1 km at the Venus distance) and 0.25 pixels (53.9 km) for AIA and XRT images, respectively. The intensity profiles we obtained have a smoothed limb due to the convolution of a step function with the instrument Point Spread Function (PSF), and are symmetrical around the middle intensity value, as expected for a symmetrical PSF. We considered the occulting radius for a given wavelength channel as the distance between this point at half-intensity and the disk center.
For each channel we obtained as many radius values as the number of selected images.
Due to the low noise, the values are invariably the same 

\begin{figure}[htbp]               
 \centering
   {\includegraphics[width=16cm]{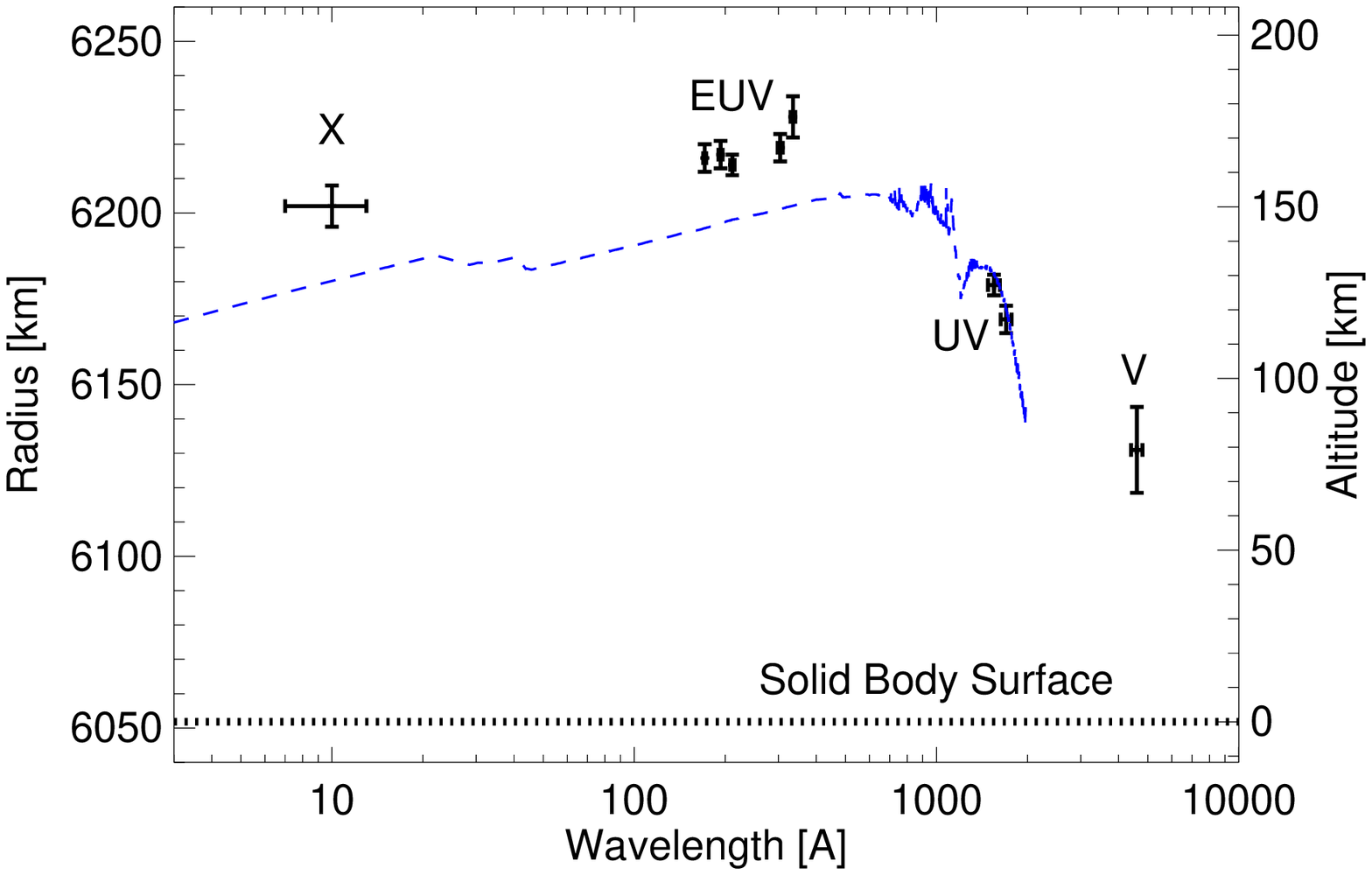}}
\caption{\footnotesize Radius of Venus measured from the transit as a function of the wavelength [data points]. The approximate FWHM of each channel \cite{Boerner2012a,Golub2007a} is marked [horizontal error bar]. Predictions from a model \cite{Fox2011a} at SZA of $90^\circ$ [blue dashed line] is shown for comparison. }
\label{fig:radius}
\end{figure}

\aftr{Due to high S/N in the optical bands, retrieved values are constant within uncertainty margin of 0.1 pixels,} corresponding to 12.6 km at the distance of Venus. 
In all the other bands we obtained normal distributions of radii, and, therefore, \aftr{the centroid position (mean value) as the best radius value and the standard deviation of the mean as its uncertainty (see Supplementary Information for details on the error analysis)}. The values are shown in Table~\ref{tab:radius} and in Fig.~\ref{fig:radius}. 

We see that the radius values follow a well-defined trend. The Venus mean solid body radius of $R_{sf} = 6051.8 \pm 1.0$ km is given by the cartographic reference system obtained from Magellan \cite{Seidelmann2002a}. 
The optical radius is in agreement with the mean cloud top altitude of $74 \pm 1$ km retrieved from Venus Express/VIRTIS \cite{Ignatiev2009a}, \aftr{and more specifically, to the expected opacity mainly due to upper haze in the first scale height above cloud tops. VEx/SOIR results \cite{Wilquet2012a} place that altitude at $73 \pm 2$ km in the 3 $\mu$m band, to be further increased by $6 \pm 1$ km to retrieve its value in the visible domain \cite{Wilquet2012a,Tanga2012a}.}
The altitude values retrieved in the EUV and X-ray bands are significantly larger by $70-100$ km than the optical radius, while the altitude in the UV band is in between ($40- 50$~km). 


\begin{table}
\caption{Observations and occulting radii and altitudes of Venus at different wavelengths during the transit of June 2012. }
\begin{center}
\begin{tabular}{lccc|lll}
\hline
Band & $\lambda$ & Start time & End time &  Radius & Altitude& Altitude\\
&&UTC&UTC&&vs clouds$^{(2)}$&vs surface\\
&[\AA] &   (5 June)   &  (6 June) &  [km]& [km] & [km] \\
\hline 
Optical & 4500 & 23:30 & 01:16  &  6131 $\pm$ 13$^{(2)}$ & 0 $\pm$ 13$^{(2)}$ & 79 $\pm$ 13$^{(2)}$\\
UV & 1700 & 22:26 & 04:14  &  6169 $\pm$ 4 & 38 $\pm$ 4& 117 $\pm$ 4\\
   &1600 & 22:26 & 04:14  &  6179 $\pm$ 3 & 48 $\pm$ 3& 127 $\pm$ 3\\
EUV & 335  & 22:23 & 04:20  &  6228 $\pm$ 6 & 97 $\pm$ 6& 176 $\pm$ 6\\
    & 304  & 22:23 & 04:20  &  6219 $\pm$ 4 & 88 $\pm$ 4& 167 $\pm$ 4\\
    & 211  & 22:23 & 04:20  &  6214 $\pm$ 3 & 83 $\pm$ 3 & 162 $\pm$ 3\\
    & 193  & 22:23 & 04:20  &  6217 $\pm$ 4 & 86 $\pm$ 4& 165 $\pm$ 4\\
    & 171  & 22:23 & 04:20  &  6216 $\pm$ 4 & 85 $\pm$ 4& 164 $\pm$ 4\\
X-rays & 10   & 22:51 & 04:24  &  6202 $\pm$ 6& 71 $\pm$ 6& 150 $\pm$ 6\\
\hline
\end{tabular}
\end{center}
{\small $^{(1)}$ ``clouds'' is here used for simplicity as reference altitude since the optical opacity in limb view is actually limited by clouds and haze on top of it without a precise boundary.

\noindent
$^{(2)}$ Upper limit for the uncertainty estimated as half of our sensitivity, i.e., 0.1 pixels, with the SDO/AIA instrument in the optical band (see Suppl. Information). }
\label{tab:radius}
\end{table} 

\aftr{The altitude that we measured in the EUV and X-ray bands corresponds to the height where the optical thickness along the tangential line of sight reaches the value $\tau = 1$. At this altitude the solar radiation is mostly absorbed by photoionization of neutral atoms, and hence this is also the altitude where a peak of the electron density is expected. In particular, the $F_1$ peak in density profiles of planetary atmospheres is associated to the absorption of EUV radiation \cite{Fox2011a}, while UV photons and soft X-rays are absorbed in slightly deeper layers. In fact, we have verified that the different Venus sizes indicated by UV and EUV data can be explained in term of tangential column densities at the retrieved altitudes (165-170 km for EUV, and 125-135 km  for UV) at terminator between the day and night hemispheres (solar zenith angle, SZA, $90^\circ$). The absorption path length, $l$, at a given altitude $h$ can be expressed as $l_i \sim \sqrt {2(R_{sf} + h)H_i}$, where $H_i$ is the atmospheric scale height of each molecular species $i$, and the optical thickness can be evaluated as $\tau(\lambda) = \sum n_i \sigma_i(\lambda) l_i$, where $n_i$ and $\sigma_i(\lambda)$ are the molecular density and absorption cross section, respectively. At 170 km CO$_2$ and CO have approximately the same mixing ratio \cite{Hedin1983a}, while at 130 km CO$_2$  dominates. A simple calculation assuming a spherical geometry shows that the neutral atmosphere is $\sim 20$ times more opaque in the EUV than in the UV at 170 km, and $\sim 80$ times at 130 km in the regions probed. }

\aftr{Fig.~\ref{fig:radius} shows a comparison between our ionospheric altitudes measured at different wavelengths with the prediction from a detailed model that includes also a dependence on the SZA \cite{Fox2011a}. There is a very good general agreement, and an accurate one at UV wavelength. However, we find that at EUV and X-ray wavelengths, although consistent with soft X-rays absorbed at slightly deeper layers than EUV \cite{Fox2011a}, the Venus atmosphere appears more opaque than expected from the model for SZA = 90$^\circ$.
This result indicates a tangential column density at heights $h \sim 150$--170 km larger than expected from the model in spherical geometry at the terminator, i.e.\ a higher neutral density than predicted by the standard model of the Venus atmosphere \cite{Hedin1983a} or a longer path. The latter might be the case if the ionosphere at large heights deviates from spherical geometry due to the pressure of the solar wind. This does not seem to be the case deeper in the atmosphere where UV radiation is absorbed.}

This is the first time that such a multi-wavelength measurement has ever been performed for a Solar System body, and it is a very rare, unplanned, opportunity we caught; in fact, the occurrence of a transit of a planet with atmosphere over the Sun, and the simultaneous availability of optical, UV, and X-ray observations will not happen again in the near future. 
This measurement is useful as a check for models of the Venus' atmosphere  \cite{Fox2007a,Witasse2006a,Patzold2007a} at the terminator, where large changes are occurring due to
the transition from sunlight to darkness \cite{Fox2011a}. It also provides an independent measurement of \aftr{the molecular column} at heights that were recently analysed by tracking data of the Venus Express Atmospheric Drag Experiment (VExADE) \cite{Rosenblatt2012a} in complement to the Neutral Mass Spectrometer (NMS) instrument onboard PVO and Magellan orbital drag measurements. On the other hand, multi-wavelength (IR, optical, X-ray) observations of transits of planets in extra-solar systems are already within the possibilities of current observing facilities on Earth and from space, but this methodology is still at its very early phases of assessment for the study of exoplanetary atmospheres. Our study of the Venus transit will be useful for planning and interpreting future data: in fact, the topic of planetary atmospheres will be certainly pursued very actively with future instrumentation (JWST, Athena, and so on), because of its relevance for understanding planet physics and life conditions in the Universe. In the mean time, our measurements are providing new information for explaining the drag that the planet atmosphere exerts on space probes currently in orbit around Venus.

\begin{addendum}
 \item We thank Jane L. Fox, Paola Testa and Paul Boerner for help and suggestions. FR, AG, GM, AM acknowledge support from  Italian Ministero dell'Universit\`a e Ricerca.
 
 \item[Author Contributions] F.R. provided the scientific leadership, supervised the work, and wrote most of the paper. A.G. made the data analysis. G.M. co-supervised the work. A.M. contributed to the analysis and interpretation of the results and to the text. T.W. provided calculations of Venus' atmosphere and text. G.P. provided important framework information. 
 
 \item[Competing Interests] The authors declare that they have no
competing financial interests.
 \item[Correspondence] Correspondence and requests for materials
should be addressed to Fabio Reale~(email: reale@astropa.unipa.it).
\end{addendum}


\end{document}